\documentclass[preprint,showpacs,preprintnumbers,superscriptaddress,
amsmath,showkeys,amssymb,aps,floatfix]{revtex4}

\usepackage{graphicx}
\usepackage{dcolumn}
\usepackage{bm}
\usepackage{longtable}
\usepackage{epsfig}
\usepackage{times}

\begin{document}

\title{Photon angular distribution and nuclear-state alignment in 
nuclear excitation by electron capture\footnote{This work is part of the 
doctoral thesis of Adriana Gagyi-P\'alffy, Giessen (D26), 2006.}}

\author{Adriana~P\'alffy}\email{Adriana-Claudia.Gagyi-Palffy@uni-giessen.de}
\affiliation{Institut f\"ur Theoretische Physik, 
Justus-Liebig-Universit\"at Giessen, Heinrich-Buff-Ring~16, 35392 
Giessen, Germany}

\author{Zolt\'an~Harman}
\author{Andrey~Surzhykov}
\author{Ulrich~D.~Jentschura}
\affiliation{Max-Planck-Institut f\"ur Kernphysik, Saupfercheckweg~1, 
69117 Heidelberg, Germany}

\date{\today}

\begin{abstract}

The alignment of nuclear states resonantly formed in nuclear excitation 
by electron capture (NEEC) is studied by means of a density matrix 
technique. The vibrational excitations of the nucleus are described by a 
collective model and the electrons are treated in a relativistic 
framework. Formulas for the angular distribution of photons emitted in 
the nuclear relaxation are derived. We present numerical results for 
alignment parameters and photon angular distributions for a number of 
heavy elements in the case of $E2$ nuclear transitions. Our results are 
intended to help future experimental attempts to discern NEEC from 
radiative recombination, which is the dominant competing process.

\end{abstract}

\maketitle


\section{Introduction}

\begin{figure}
\begin{center}
\includegraphics[width=0.9\textwidth]{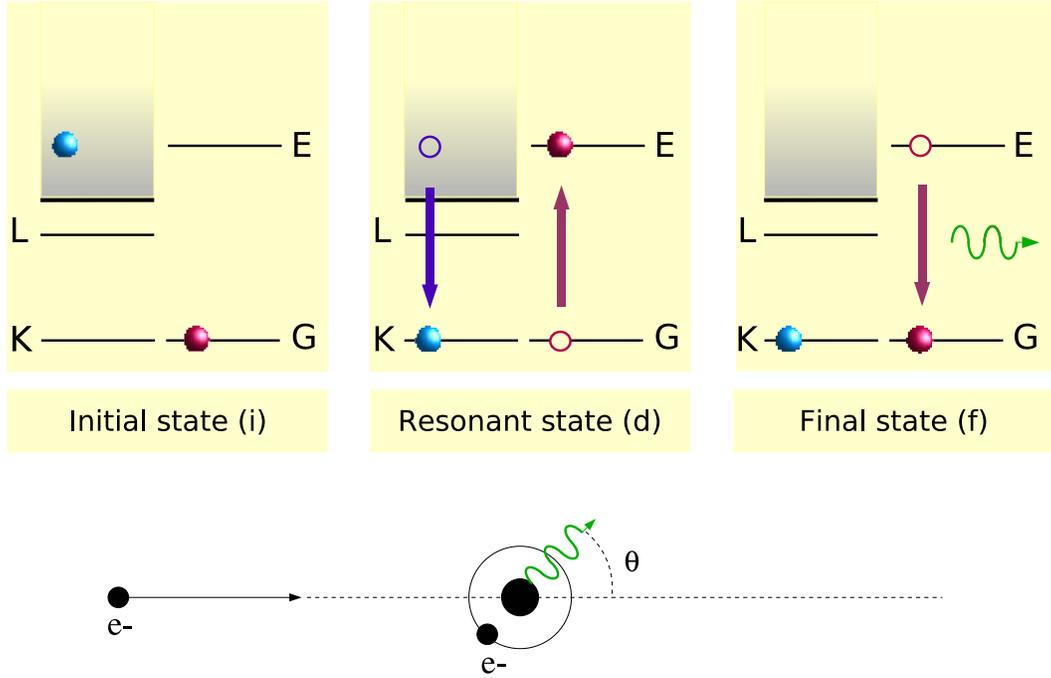}
\caption{\label{neecplot} NEEC recombination mechanism of a continuum 
electron into the $K$ shell of a bare ion. The nucleus is schematically 
represented as undergoing the transition from the ground state (G) to 
the excited state (E) and again to its ground state. The figure below 
shows the geometry of the process.}
\end{center}
\end{figure}

The process studied in this paper consists of two steps. First, a free 
electron is recombined into an electronic shell of a positive,  
preferably highly charged ion while resonantly transferring its 
excess energy to the nucleus. This step is referred to as nuclear 
excitation by electron capture (NEEC) in the literature. It is the 
time-reversed process of internal conversion (IC) and can also 
be regarded as the nuclear analogue of dielectronic recombination (DR),
where a bound electron is resonantly excited. In the second step, 
the excited nuclear state thus formed decays radiatively. The whole 
process is illustrated in Fig.~\ref{neecplot}. Proposed for the first 
time in Ref.~\cite{Goldanskii}, the NEEC recombination mechanism has not 
been observed experimentally yet.

Related processes at the borderline between nuclear and atomic physics 
have been experimentally confirmed, such as nuclear excitation by 
electronic transition (NEET), that is a decay mode of the electron 
atomic shell in which energy is transferred to the nucleus. NEET has 
been observed in ${}^{189}$Os~\cite{Oto77}, ${}^{237}$Np~\cite{Sai90}, 
and, most recently, in ${}^{197}$Au~\cite{Kishimoto}. For its 
time-reversed process, the bound internal conversion (BIC), a direct 
experimental evidence was not found until recently~\cite{Carreyre}.

The detection of the recombined ions or the emitted radiation 
characteristic for the nuclear decay can be used to observe 
NEEC experimentally. The position of the peak in the 
recombination cross section or the spectrum of emitted photons 
directly yields the nuclear transition energy. The detected spectra may 
provide further information on nuclear decay mechanisms and level 
population. Thus, the study of NEEC in recombination experiments can
also be regarded as a possible novel tool to investigate nuclear 
properties.

In a previous paper~\cite{neec}, we presented total NEEC cross 
sections for a range of heavy isotopes. Recently, the quantum
interference of NEEC and the background process of radiative recombination (RR)
was investigated~\cite{interf}. In the present work we study a 
new aspect, namely, the alignment, i.e., the magnetic sublevel 
population of the nuclear excited states generated in the resonant 
recombination process. 
The alignment of the nuclear excited state also allows one 
to derive the angular distribution of the radiation emitted in the 
nuclear relaxation process and the cross section differential with 
respect to the photon emission angle.
Differential cross sections for the similar process of DR have been
investigated theoretically~\cite{Chen95,Gail98,Zakowicz,Zakowicz2} and
experimentally~\cite{Kandler95,Ma}.

The photon angular distribution is 
of general interest for experimental implementations aiming at observing 
NEEC by the detection of radiation. Furthermore, as found in this work, 
the photon emission pattern is substantially different from the one of 
the concurrent process of RR, which may help 
to discern the two channels.

The reorientation of the nuclear axis caused by 
the electric field of a charged particle that excites the nucleus 
is a well-studied process in nuclear physics. The 
change in the nuclear spin directions following the Coulomb excitation 
of the nucleus by collisions with low-energy charged particles affects 
the angular distribution of the emitted $\gamma$ rays \cite{Breit}.  In 
the case of the electric radiative transitions, NEEC can be regarded as 
a Coulomb excitation with free electrons that are in the end recombined into 
a bound state of the ion.

We apply a density matrix formalism to describe the alignment of 
rotationally excited nuclear states formed by NEEC and the angular 
distribution of the de-excitation radiation. Nuclear states are treated 
in the framework of the nuclear collective model and the electrons are 
described by Dirac four-component wave functions, as necessary for the 
heavy elements studied here. The electron-nucleus interaction operator 
is expanded in spherical multipoles to facilitate the evaluation of transition matrix
elements. This theoretical approach is 
presented in detail in Sec.~\ref{theory}. NEEC nuclear alignment 
parameters are presented for the case of $E2$ excitations in heavy 
highly charged ions and differential cross sections are compared to the 
same quantity corresponding to the background process of RR in 
Sec.~\ref{results}. The relevance of these results for future 
experiments targeting the observation of NEEC is discussed. We briefly 
summarize the findings of the paper in Sec.~\ref{summary}. Atomic units are used throughout this work unless otherwise specified.


\section{\label{theory} Theoretical formalism}


Since its introduction in 1927 by von Neumann and Landau, the density 
matrix approach has been found to be a useful and elegant tool in many fields 
of modern physics. For applications in atomic physics, and combined 
particularly with the concept of spherical tensors, this approach was 
originally developed by Fano~\cite{FaR59} in the 1950s. Since 
then, the density matrix theory has been utilized successfully in  
many case studies on atomic collisions, e.g., for describing the excitation of 
atomic autoionizing states, polarization effects in radiative and 
Auger decays, cascade processes, or lifetime interferences in resonantly excited atoms.

In this paper we apply a density matrix formalism to describe the 
two--step process in which (i) a free (or quasi--free) electron is 
captured into the bound state of an initially bare ion with excitation 
of the atomic nucleus which (ii) subsequently decays under the emission 
of characteristic radiation. Because the properties of this radiation 
are closely related to the alignment of the excited nuclear state, we 
first have to investigate the population of these states as it arises 
due to the electron capture process. Therefore, in Section~\ref{theory_alignment}, 
we derive the general formulas for the density matrix of the 
excited nuclear states produced by electron capture. In particular, 
we show how the sublevel population of these states can be described in 
terms of the so--called alignment parameters. The calculation of these 
parameters involves the NEEC transition amplitudes and requires the use 
of a nuclear model. Following Ref.~\cite{neec}, the 
NEEC transition amplitudes are derived in 
Section~\ref{theory_calculation_amplitude}. Finally, in 
Section~\ref{theory_decay} we consider the subsequent nuclear decay and 
obtain the angular distribution of the de--excitation photons with the
help of the alignment parameters.

\subsection{Alignment of the excited nuclear state}
\label{theory_alignment}

Within the density matrix theory, the state of a physical system is 
described in terms of statistical (or density) operators. 
These operators can be considered to represent, for instance, an 
ensemble of systems which are---altogether---in either a pure quantum 
state or in a mixture of different states with any degree of coherence. 
Then, the basic idea of the density matrix formalism is to accompany 
such an ensemble through the collision process, starting from a 
well-defined initial state and by passing through one or, possibly, 
several intermediate states until the final state of the collision 
process is attained.
 
In NEEC, the initial state of the 
combined system is given by the electron with a well--defined 
asymptotic momentum $\bm{p}$ and spin projection $m_s$, and an ion which 
is specified in term of its nuclear charge $Z$ and its initial nuclear spin 
quantum numbers $I_i$ and projection $M_i$. (We note here that we consider
recombination into bare or closed shell ions, thus the angular momentum
of the bound electron shell is zero.) Assuming that these two 
subsystems, ion and electron, are uncorrelated, the overall initial 
density operator is given as the direct product of the two initial 
subsystems' density operators:
\begin{equation}
    \label{rho_initial_general}
    {\rho}_i = {\rho}_{ion} \otimes {\rho}_{e} \, . 
\end{equation}
If neither the electrons nor the ions are polarized initially, 
an averaging over the magnetic quantum numbers can be performed and
the tensor product can be written as
\begin{equation}
    \label{rho_initial_unpolarized}
    {\rho}_i = \frac{1}{2} \frac{1}{2 I_i + 1} \, 
    \sum\limits_{m_s M_i} |{\bm{p} m_s}\rangle |N_i I_i M_i\rangle
    \langle{N_i I_i M_i}| \langle{\bm{p} m_s}| \, , 
\end{equation}
where $N_i$ denotes all the additional quantum numbers needed for a
unique specification of the nuclear states. 

In the intermediate state $d$ formed by the capture of the electron, the 
statistical operators have to describe both the electron in some bound 
ionic state $|{n_d \kappa_d j_d m_d}\rangle$ as well as the state of the 
excited nucleus $|{N^*_d I_d M_d}\rangle$. Here, $n_d$, $\kappa_d$, $j_d$ and 
$m_d$ are the principal quantum number, Dirac angular momentum quantum number, total angular momentum quantum number  
and magnetic quantum number of the bound one-electron state, 
respectively.  As known from density matrix theory, the statistical 
operators of the initial and the (subsequent) intermediate states of the 
system are connected by
\begin{equation}
   \label{rho_intermediate_general}
   {\rho}_d = {T}_{en} {\rho}_i {T}_{en}^\dagger \, , 
\end{equation}
where ${T}_{en}$ is the transition operator for the electron--nucleus 
interaction which causes the nuclear excitation. The particular 
form of this operator will be given in Section~\ref{theory_calculation_amplitude}.

Instead of applying equation (\ref{rho_intermediate_general}), in 
practice, it is often more convenient to rewrite the statistical 
operators in a matrix representation. For instance, in a basis with 
well-defined angular momenta, the intermediate-state density matrix is 
given by
\begin{eqnarray}
   \label{matrix_intermediate_general}
  && \left\langle{N_d^* I_d M_d,\, n_d \kappa_d j_d m_d  }
  \left\vert{{\rho}_d}\right
  \vert{N_d^* I_d M'_d,\, n_d \kappa_d j_d m'_d}\right\rangle  = \nonumber \\
&& \frac{1}{2} \frac{1}{2 I_i + 1} \,
   \sum\limits_{m_s M_i} 
   \left\langle{N_d^* I_d M_d,\, n_d \kappa_d j_d m_d }\left
   \vert{{T}_{en}}\right\vert{N_i I_i M_i,\, \bm{p} m_s }\right\rangle
   \nonumber \\
   && \times  
   \left\langle{N_d^* I_d M'_d,\, n_d \kappa_d j_d m'_d }\left
   \vert{{T}_{en}}\right\vert{N_i I_i M_i,\, \bm{p} m_s }\right\rangle^*
   \, ,
\end{eqnarray}
assuming that both the incident electrons and ions are initially 
unpolarized [see Eq.~(\ref{rho_initial_unpolarized})]. Indeed, the 
intermediate--state density matrix~(\ref{matrix_intermediate_general}) 
still contains the complete information about the NEEC process and, 
thus, can be used to derive all the properties of the bound electron and 
the excited nucleus. For instance, assuming that the magnetic states 
$m_d$ of the bound electron remain unobserved in the particular 
experiment, we may characterize the sublevel population of the excited 
nucleus $|{N_d^* I_d}\rangle$ in terms of the nuclear density matrix
\begin{eqnarray}
   \label{matrix_ion_general}
   \left\langle{N_d^* I_d M_d}\left\vert{{\rho}^{ nucl}_d}
   \right\vert{N_d^* I_d M'_d} \right\rangle
   & = & \frac{1}{2} \frac{1}{2 I_i + 1} \,
   \sum\limits_{m_s M_i m_d} 
   \left\langle{N_d^* I_d M_d,\, n_d \kappa_d j_d m_d }\left
   \vert{{T}_{en}}\right\vert{N_i I_i M_i,\, \bm{p} m_s}\right\rangle
   \nonumber \\
   & \times & 
   \left\langle{ N_d^* I_d M'_d,\, n_d \kappa_d j_d m_d}\left
   \vert{{T}_{en}}\right\vert{N_i I_i M_i,\, \bm{p} m_s}\right\rangle^* \, ,
\end{eqnarray}
which is obtained from Eq.~(\ref{matrix_intermediate_general}) by taking 
the trace over all unobserved quantum numbers of the electron.

As seen from Eq.~(\ref{matrix_ion_general}), the information about the 
states of the excited nucleus produced by the electron capture into 
the ion is now contained in the transition matrix elements 
$\left\langle{ N_d^* I_d M_d,\, n_d \kappa_d j_d 
m_d}\left\vert{{T}_{en}}\right\vert{N_i I_i M_i,\, \bm{p} m_s 
}\right\rangle$. These matrix elements contain the wave function $|{ 
\bm{p} m_s}\rangle$ of a free electron with a definite asymptotic 
momentum. For further simplification of the intermediate nuclear 
spin--density matrix, it is therefore necessary to decompose this 
continuum wave into partial waves $|{\epsilon \kappa j m}\rangle$, in 
order to apply later the standard techniques from the theory of angular 
momentum. As discussed previously~\cite{Eichler}, however, special care 
has to be taken about the choice of the quantization axis since this 
directly influences the particular form of the partial wave 
decomposition. Using, for example, the direction of the electron 
momentum $\bm{p}$ as the quantization axis, the full expansion of the 
continuum wave function is given by~\cite{Eichler}
\begin{equation}
   \label{electron_decomposition}
   |{\bm{p} m_s}\rangle = \sum\limits_{\kappa} i^l {\rm e}^{i \Delta_\kappa} \,
   \sqrt{4 \pi (2l + 1)} \,
   \left\langle{l 0\  1/2 m_s}\left\vert\right.{j m_s}\right\rangle \, |{\epsilon \kappa j m_s}\rangle \, ,
\end{equation}
where the summation runs over all partial waves $\kappa = \pm 1, \pm 2 
$, \ldots, along all values of Dirac's angular momentum quantum number 
$\kappa = \pm (j + 1/2)$ for $l = j \pm 1/2$. The symbol
$\left\langle{j_1 m_1\  j_2 m_2}\left\vert\right.{j_{12} m_{12}}\right\rangle$
generally represents the Clebsch-Gordan coefficients for the coupling of two angular
momenta $j_1$ and $j_2$ to $j_{12}$. In our notation, the 
orbital momentum $l$ represents the parity $(-1)^l$ of the 
partial waves $|{\epsilon \kappa j m_s}\rangle$, and $\Delta_\kappa$ is 
the Coulomb phase shift given by~\cite{Eichler}
\begin{equation}
\Delta_{\kappa}=\frac{1}{2}\ {\rm arg}
\left(\frac{-\kappa+i\nu/W}{s+i\nu}\right)
-{\rm arg}(\Gamma(s+i\nu))+\frac{\pi(l+1-s)}{2}\ ,
\label{C-phase}
\end{equation}
with $W=E\alpha^2$, $\nu=\alpha ZW/\sqrt{W^2-1}$, 
$s=\sqrt{\kappa^2-(\alpha Z)^2}$. Here, $\alpha$ is the fine-structure 
constant and $E$ is the total electron 
energy. In the case of capture into ions with an initially closed shell -- 
i.e., He-like -- configuration, the phases can be approximated by using 
an effective nuclear charge of $Z_{\rm eff} = Z - N_b$ in 
Eq.~(\ref{C-phase}), with $N_b$ being the number of bound electrons in 
the ion. The sufficiency of this approximation is confirmed by 
calculating the electrostatic potential induced by the screening 
electrons in the Dirac-Fock approximation~\cite{Dya89} and numerically determining 
the phases for the combined nuclear and screening potentials.

Using the decomposition (\ref{electron_decomposition}) of the continuum 
wave function, the intermediate nuclear density matrix 
(\ref{matrix_ion_general}) can be rewritten in the form
\begin{eqnarray}
   \label{matrix_ion_final}
   \left\langle{N_d^* I_d M_d}\left\vert{{\rho}^{nucl}_d}
   \right\vert{N_d^* I_d M'_d} \right\rangle
   & = & \frac{1}{2} \frac{4 \pi}{2 I_i + 1} \,
   \sum\limits_{m_s M_i m_d} \sum\limits_{\kappa \kappa'}
   i^{l - l'} \, {\rm e}^{i (\Delta_\kappa - \Delta_{\kappa'})} \,
   \sqrt{(2l + 1)(2l' + 1)} \,
   \\
   & \times & \left\langle{l 0 \ 1/2 m_s}\left\vert\right.{j m_s}\right\rangle    
   \left\langle{N_d^* I_d M_d,\, n_d \kappa_d j_d m_d }\left\vert{{T}_{en}}
   \right\vert{N_i I_i M_i,\, \epsilon \kappa j m_s}\right\rangle
   \nonumber \\
   & \times & \left\langle{l' 0 \ 1/2 m_s}\left\vert\right.{j' m_s}\right\rangle 
   \left\langle{N_d^* I_d M'_d,\, n_d \kappa_d j_d m_d}\left\vert{{T}_{en}}
   \right\vert{N_i I_i M_i,\, \epsilon \kappa' j' m_s }\right\rangle^* \,.
   \nonumber 
\end{eqnarray}
Equation (\ref{matrix_ion_final}) represents the most general form of the 
intermediate nuclear density matrix which can now be used to study the 
properties of the excited nucleus $|{N_d^* I_d}\rangle$ following the 
capture of a free electron. For the analysis of the 
radiative de-excitation of such a nucleus, however, it is more 
convenient to represent its intermediate state in terms of the 
statistical tensors of rank $k$
\begin{equation}
   \label{stat_tensors_definition}
   \rho_{kq}(N_d^* I_d) = \sum\limits_{M_d M'_d} (-1)^{I_d - M_d'} \,
   \left\langle{I_d M_d \ I_d -M_d'}\left\vert\right.{k q}\right\rangle \, 
   \left\langle{N_d^* I_d M_d}\left\vert{{\rho}^{nucl}_d}
   \right\vert{N_d^* I_d M'_d} \right\rangle \,.
\end{equation}
Although both the density matrix~(\ref{matrix_ion_final}) and the statistical 
tensors~(\ref{stat_tensors_definition}) contain the same physical information, 
the latter form enables one to exploit the rotational symmetry of free 
atoms and ions. By inserting the density matrix~(\ref{matrix_ion_final}) 
into the definition~(\ref{stat_tensors_definition}), we finally 
obtain the statistical tensors of the intermediate state as
\begin{eqnarray}
   \label{stat_tensor_excited_state}
   \rho_{kq}(N^*_d I_d) & = & \frac{4 \pi}{2 (2I_i + 1)} \,
   \sum\limits_{m_{s} M_{i} m_d} \, 
   \sum\limits_{\kappa \kappa'} \,
   \sum\limits_{M_{d}  M'_{d}}  i^{l - l'} \,
   {\rm e}^{i (\Delta_{\kappa} - \Delta_{\kappa'})} \,\sqrt{(2l+1)(2l'+1)} \,
   (-1)^{I_d - M'_{d}} 
   \nonumber \\[0.2cm]
   &\times& \left\langle{l 0 \ 1/2 m_{s}}\left\vert\right.{j m_{s}}\right\rangle \,
   \left\langle{l' 0 \ 1/2 m_{s}}\left\vert\right.{j' m_{s}}\right\rangle \,
   \left\langle{I_d M_{d}\ I_d -M'_{d}}\left\vert\right.{k q}
   \right\rangle \nonumber \\[0.2cm]
   &\times& 
   \left\langle{N^*_d  I_d  M_{d}, \,  n_d \kappa_d j_d  m_d}\left\vert{{T}_{en}}
   \right\vert{N_i  I_i  M_{i}, \epsilon \kappa j m_s}\right\rangle   
   \nonumber \\[0.2cm]
   &\times& 
   \left\langle{N^*_d  I_d  M'_{d},\,  n_d  \kappa_d j_d  m_d}\left\vert{{T}_{en}}
   \right\vert{N_i  I_i  M_{i}, \,  \epsilon \kappa' j' m_s}\right\rangle^* \, .
\end{eqnarray}

Further evaluation of these statistical tensors within the framework of 
the nuclear collective model is discussed in the next section. The spin 
state of the excited nucleus is described by the reduced statistical 
tensors or alignment parameters
\begin{equation}
\mathcal{A}_k(N^*_dI_d)=\frac{\rho_{k0}(N^*_dI_d)}{\rho_{00}(N^*_dI_d)}\ ,
\label{A_s}
\end{equation}
which are directly related to the cross section
for the population of the different nuclear magnetic 
sublevels $|N^*_dI_dM_d\rangle$.

\subsection{Calculation of the transition amplitudes}
\label{theory_calculation_amplitude}

The expression of the statistical tensors involves the matrix element of 
the transition operator ${T}_{en}$ that describes the electron-nucleus 
interaction in the case of electric transitions of the nucleus. We adopt 
in the following the Coulomb gauge as it allows the separation of the 
dominant Coulomb interaction between the electronic and nuclear degrees 
of freedom. The transition operator corresponds in lowest order to the 
interaction Hamiltonian $H_{en}$ applied in~\cite{neec}:
\begin{equation}
{T}_{en}=\int d^3r_n \frac{\rho_n(\vec{r}_n)}{|\vec{r}_e-\vec{r}_n|}\ .
\label{T_op}
\end{equation}
In the above equation, $\rho_n(\vec{r}_n)$ is the nuclear charge 
density, $\vec{r}_n$ denotes the nuclear coordinate and $\vec{r}_e$ the 
electronic coordinate. The integration is performed over the whole 
nuclear volume. The matrix element of the transition operator ${T}_{en}$ 
enters the expression of the NEEC rate for electric transitions 
\cite{neec}
\begin{eqnarray}
&&Y_n^{i \to d}=\frac{2\pi}{2(2I_i+1)}\sum_{{M_{i}} m_s}\sum_{M_{d} m_d}
\nonumber \\
&&
\int d\Omega_{p}
\left\vert\left\langle N^*_dI_dM_d,\, n_d\kappa_d j_d m_d
\left\vert{T}_{en}\right\vert N_iI_iM_{i},\, \bm{p}m_s\right\rangle\right\vert^2\rho_i\,,
\label{Yrate}
\end{eqnarray}
where $\rho_i$ is the density of the initial electronic states and $\int \Omega_p$
denotes integration over all possible directions of the incoming electron.

In order to describe the nuclear transition we use a 
phenomenological collective model~\cite{Greiner} that interprets the 
characteristic band structures in the energy range up to 2 MeV in the 
case of deformed even-even nuclei as vibrations and rotations of the 
nuclear surface. The even-even nuclei have usually a low-lying $2^+$ 
first excited state, which is characterized by a strong electric $E2$ 
transition to the ground state.  The nuclear surface is parametrized as
\begin{equation}
R(\theta_n,\varphi_n)=R_0\Big(1+\sum_{\ell=0}^\infty
\sum_{m=-\ell}^{\ell} \alpha_{\ell m}^*Y_{\ell m}(\theta_n,\varphi_n)\Big),
\label{param}
\end{equation}
where the amplitudes $\alpha_{\ell m}$ describe the 
deviations of the nuclear surface with respect to the sphere of radius 
$R_0$ and thus serve as collective coordinates. Using this 
parametrization and requiring that the charge be homogeneously 
distributed, the nuclear charge density can be written as
\begin{equation}  
\rho_n(\vec{r}_n)=\rho_0\Theta\left(R(\theta_n,\varphi_n)-r_n\right)\ ,
\end{equation}
with the constant charge density of the undeformed nucleus given by 
$\rho_0=\frac{3Z}{4\pi R_0^3}$. Performing a Taylor 
expansion of the Heaviside function $\Theta(R-r_n)$ around $R_0$ we obtain
\begin{equation}
\rho_n (\vec{r}_n)=\rho_0 \Theta(R_0-r_n)+\rho_0\delta(r_n-R_0)R_0\sum_{\ell m}\alpha_{\ell m}^*
Y_{\ell m}(\theta_n,\varphi_n)+\ldots\ .
\end{equation} 
As the vibration amplitudes of the nuclear surface are supposed to be 
small, we neglect the terms of higher order in the collective 
coordinates $\alpha_{\ell m}$. While the first term $\rho_0 
\Theta(R_0-r_n)$ in the above equation corresponds 
to a round nucleus in its ground state, the second term is characterizing 
the nuclear excitation and enters the expression of the transition 
operator ${T}_{en}$ in Eq.~(\ref{T_op})~\cite{neec}. We can therefore write the 
transition operator as
\begin{equation}
{T}_{en}=\rho_0R_0\sum_{\ell m}\alpha_{\ell m}^*\int d^3r_n
\frac{\delta(r_n-R_0)Y_{\ell m}(\theta_n,\varphi_n)}{|\vec{r}_e-\vec{r}_n|}\ .
\end{equation}
It is more convenient to express the collective coordinates 
$\alpha_{\ell m}$ in terms of the spherical components of the electric 
multipole transition moment $Q_{\ell m}$, defined as~\cite{Ring}
\begin{equation}
Q_{\ell m}=\int d^3r r^{\ell} Y_{\ell m}(\theta,\varphi)\rho_n(\vec{r})\ .
\end{equation}
The interaction that accounts for the electric transitions of the 
nucleus then yields
\begin{equation}
{T}_{en}=\sum_{\ell m}\frac{Q_{\ell m}}{R_0^{\ell}}\int d^3r_n
\frac{\delta(r_n-R_0)Y_{\ell m}^*(\theta_n,\varphi_n)}{|\vec{r}_e-\vec{r}_n|}\ .
\label{H_en:work}
\end{equation} 
In the calculation of the matrix element, we use the multipole expansion
\begin{equation}
\frac{1}{|\vec{r}_e-\vec{r}_n|}=\sum_{L=0}^{\infty}\sum_{M=-L}^L 
\frac{4\pi}{2L+1}Y_{LM}(\theta_n,\varphi_n)Y^*_{LM}(\theta_e,\varphi_e)
\frac{r^L_{<}}{r^{L+1}_{>}}\,,
\end{equation}
where $r_<$ and $r_>$ stand for the smaller and the larger 
of the two radii $r_e$ and $r_n$, respectively. The integration over the 
nuclear angular coordinates brings us to the following expression for the 
Coulomb interaction:
\begin{equation}
{T}_{en}=\sum_{LM}\frac{4\pi}{2L+1}\frac{Q_{LM}}{R_0^L}
Y^*_{LM}(\theta_e,\varphi_e)\int_0^{\infty}dr_nr_n^2
\frac{r^L_{<}}{r^{L+1}_{>}}\,\delta(r_n-R_0)\,.
\end{equation} 
The matrix element of the transition operator reads
\begin{eqnarray}
&&\left\langle{N^*_d  I_d  M_{d}, \,  n_d  \kappa_d j_d  m_d}
\left\vert{{T}_{en}}\right\vert{
N_i  I_i  M_{i},\,  \epsilon \kappa j m_s}\right\rangle
= \nonumber \\
&&
\frac{1}{R_0^L}\sum_{LM}\frac{4\pi}{2L+1}\left\langle N^*_d I_d M_d\left\vert 
Q_{LM}\right\vert N_i I_i M_i\right\rangle
\nonumber \\ &&\times
\left\langle n_d \kappa_d j_d m_d\left\vert 
Y^*_{LM}(\theta_e,\varphi_e)\int_0^{\infty}dr_nr_n^2\frac{r^L_{<}}{r^{L+1}_{>}}
\delta(r_n-R_0)\right\vert \epsilon \kappa j m_s\right\rangle\ .
\label{elme1}
\end{eqnarray}
We write the matrix element of the electron-nucleus interaction operator 
as a product of the nuclear and electronic parts. It is more convenient 
to use the reduced matrix element of the electric multipole operator 
$Q_{LM}$, defined as \cite{Edmonds}
\begin{eqnarray}
\left\langle N^*_dI_dM_d\left\vert Q_{LM}\right\vert N_iI_iM_i\right\rangle &=&
\frac{(-1)^{I_i-M_i}}{\sqrt{2L+1}}
\left\langle{I_dM_d\  I_i-M_i}\left\vert\right.{L M}\right\rangle \nonumber \\
&\times&\left\langle N^*_dI_d\left\Vert Q_{L}\right\Vert N_i I_i\right\rangle\,.
\end{eqnarray}
The modulus square of the reduced matrix element of the electric multipole operator
can be related to the reduced electric ($E$) transition 
probability of a certain multipolarity $L$,
\begin{equation}
B (EL,I_i\to  I_d)=\frac{1}{2I_i+1}|\langle N^*_d I_d\|Q_L\|N_iI_i\rangle |^2\ ,
\end{equation}
whose value can be taken from experimental results. For a given 
multipolarity $L$, the matrix element can be written as
\begin{eqnarray}
&&\left\langle N^*_d I_d M_{d},\, n_d \kappa_d j_d m_d\left\vert{T}_{en}
\right\vert N_i I_i  M_{i},\, \epsilon \kappa j m_s\right\rangle
=
\nonumber \\
&&\sum_{\mu=-L}^L(-1)^{I_d+M_{i}+L+\mu+m_s+3j_d}R_0^{-(L+2)} R_{L,\kappa_d,\kappa} 
\left\langle N^*_d I_d\left\Vert Q_L\right\Vert N_iI_i\right\rangle
\nonumber \\ &&\times
\sqrt{2j_d+1}\sqrt{\frac{4\pi}{(2L+1)^3}}\, 
\left\langle{I_i-M_i\ I_dM_d}\left\vert\right.{ L \mu}\right\rangle
\nonumber \\ 
&&\times \,
  \left\langle{j-m_s\ j_dm_d}\left\vert\right.{ L-\mu}\right\rangle\,
\left\langle{j_d1/2\ L0}\left\vert\right.{ j1/2}\right\rangle\ , 
\label{transition_amplitude_general}
\end{eqnarray}
with the electronic radial integral $R_{L,\kappa_d,\kappa}$ given by
\begin{eqnarray}
&&R_{L,\kappa_d,\kappa}=\frac{1}{R_0^{L-1}}\int_0^{R_0} dr_e r_e^{L+2}
\bigg(f_{n_d\kappa_d}(r_e)f_{\epsilon\kappa}(r_e)+
g_{n_d\kappa_d}(r_e)g_{\epsilon\kappa}(r_e)\bigg) \nonumber \\
&&+R_0^{L+2}\int_{R_0}^\infty dr_e r_e^{-L+1}
\bigg(f_{n_d\kappa_d}(r_e)f_{\epsilon\kappa}(r_e)+
g_{n_d\kappa_d}(r_e)g_{\epsilon\kappa}(r_e)\bigg)\ .
\label{rrs}
\end{eqnarray}
In the electronic radial integrals, $g_{\epsilon\kappa}$ and 
$f_{\epsilon\kappa}$ are the large and small radial components of the 
relativistic partial continuum electron wave function
\begin{equation}
\phi_{\epsilon\kappa j m_s}(\vec{r}_e) = 
\langle \vec{r}_e \, |\epsilon\kappa jm_s\rangle=\left(\begin{array}{c} g_{\epsilon\kappa}(r_e)
\Omega_{\kappa}^{m_s}(\theta_e,\varphi_e)\\if_{\epsilon\kappa}(r_e)
\Omega_{-\kappa}^{m_s}(\theta_e,\varphi_e)\end{array}
\right)\ ,
\label{cont_wf}
\end{equation}
and $g_{n_d\kappa_d}$ and $f_{n_d\kappa_d}$ are the components of 
the bound Dirac wave functions
\begin{equation}
\phi_{n_d\kappa_d j_d m_d}(\vec{r}_e) = 
\langle \vec{r}_e \, |n_d\kappa_d j_dm_d\rangle=\left(
\begin{array}{c} g_{n_d\kappa_d}(r_e)
\Omega_{\kappa_d}^{m_d}(\theta_e,\varphi_e)\\
if_{n_d\kappa_d}(r_e)\Omega_{-\kappa_d}^{m_d}(\theta_e,\varphi_e)\end{array}
\right)\ ,
\label{bound_wf}
\end{equation}
with the spherical spinor functions $\Omega_{\kappa}^{m}$. The radial 
integral $R_{L,\kappa_d,\kappa}$ is calculated numerically.  For 
the particular case of the $0^+\rightarrow 2^+$ $E2$ transitions, the 
transition amplitude reads
\begin{eqnarray}
&&\left\langle N^*_d2M_d,\, n_d\kappa_d j_d m_d\left\vert{T}_{en}
\right\vert N_i00,\, \epsilon \kappa jm_s\right\rangle=
\frac{\sqrt{4\pi}}{\sqrt{125}}R_0^{-4}(-1)^{M_d+m_s+3j_d}\sqrt{2j_d+1}
\nonumber \\
&&\times
\left\langle N^*_d2\left\Vert Q_2\right\Vert N_i0\right\rangle\ 
\left\langle{ j-m_s\  j_dm_d}\left\vert\right.{ 2-M_d}\right\rangle\, 
\left\langle{ j_d1/2\  20}\left\vert\right.{ j1/2}\right\rangle
R_{2,\kappa_d,\kappa} \,.
\label{me}
\end{eqnarray}
The Clebsch-Gordan coefficient $\left\langle{ j-m_s\ 
j_dm_d}\left\vert\right.{ 2-M_d}\right\rangle $ imposes that 
$M_d=m_s-m_d$, therefore $M_d=M_d'$ and $q=0$ in 
Eq.~(\ref{stat_tensor_excited_state}).

\subsection{Radiative de-excitation of the nucleus}
\label{theory_decay}

By making use of the multipole transition amplitudes 
(\ref{transition_amplitude_general}), we are able to calculate now the 
alignment parameters (\ref{A_s}) of the nucleus excited by 
electron capture. A detailed knowledge of the alignment parameters is 
required for the analysis of the subsequent de-excitation of the 
nucleus which may result in the emission of one (or several) photons 
until the nuclear ground state is reached. This photon emission is 
characterized (apart from its known energy) by its angular distribution 
and polarization. The relations of both of these properties to the 
alignment $\mathcal{A}_k$ of the excited nuclear state are well known 
since the early 1960s and have been discussed in detail elsewhere 
\cite{Fer65, Sie65, RoB67, Bal84}. In the present work, therefore, we 
will restrict ourselves to a rather short account of the basic formulas. 
For instance, the angular distribution of the gamma rays emitted in the 
transition from the exited state $|N^*_d I_d M_d \rangle$ to the nuclear
ground state $|N_f I_f M_f \rangle$ is given by
\begin{equation}
   \label{angular_distribution_general}
   \frac{d\sigma_{\rm NEEC}}{d\Omega}(\theta)  = 
   \frac{\sigma_{\rm NEEC}}{4\pi}
   \left( 1+\sum_{k=2,4,\ldots}f_k(N^*_dI_d,N_fI_f) 
   \mathcal{A}_k(N^*_dI_d)P_k(\cos\theta)\right)\,,
\end{equation}
where $\sigma_{\rm NEEC}$ is the total cross section for NEEC followed 
by the radiative decay of the excited nucleus and $\theta$ denotes the 
angle of the photons with respect to the momentum $\bm{p}$ of the 
incoming electrons (chosen as the $z$-axis). As seen from 
Eq.~(\ref{angular_distribution_general}), the angular dependence of the 
photon emission results from the Legendre polynomials $P_k({\rm 
cos}\,\theta)$ which are weighted by the alignment parameters 
$\mathcal{A}_k(N^*_dI_d)$ and the coefficients $f_k(N^*_dI_d,N_fI_f)$. 
In contrast to the alignment parameters, these geometrical coefficients are 
independent on the nuclear excitation process and just account for the 
initial and the final nuclear states \cite{RoB67,Bal84,Andrey_geo},
\begin{eqnarray}
   \label{structure_function}
   f_k(N^*_dI_d,N_fI_f) & = & 
   \frac{\sqrt{2I_d+1}}{2}\sum_{L L' p p'}i^{L'+ p'-L-p}
   (-1)^{I_f+I_d+k+1} \\ &\times&\sqrt{(2L+1)(2L'+1)}\,
   \left\langle{L1\ L'-1}\left\vert\right.{k0}\right\rangle \nonumber \\
   &\times&
   \left(1+(-1)^{L+p+L'+p'-k}\right)\left\{\begin{array}{ccc} 
   L&L'&k \\I_d&I_d&I_f\end{array}\right\}
   \nonumber \\ &\times&
   \frac{\left\langle N^*_d I_d\left\Vert{T}_{nr}(L,p)\right\Vert 
   N_f I_f\right\rangle^* \left\langle N^*_d I_d\left\Vert{T}_{nr}(L',p')\right\Vert 
   N_f I_f\right\rangle}{\sum_{Lp}\left\vert\left\langle N^*_d I_d\left\Vert
   {T}_{nr}(L,p)\right\Vert 
   N_f I_f\right\rangle\right\vert^2}\,.\nonumber
\end{eqnarray}
Here, $\left\langle N^*_d I_d\left\Vert{T}_{nr}(L,p)\right\Vert N_f 
I_f\right\rangle$ denotes the reduced matrix element for the $|N^*_d I_d 
M_d \rangle \rightarrow |N_f I_f M_f \rangle$ decay under the 
emission of a photon with angular momentum (or 
multiplicity) $L$ and parity $(-1)^{p + L}$. The form of the 
nucleus-radiation field multipole interaction operator 
${T}_{nr}(L,p)$ is given in, e.g., Refs.~\cite{interf,Ring}.

Equations (\ref{angular_distribution_general}) and 
(\ref{structure_function}) display the general form of the photon 
angular distribution for the decay of an aligned system \cite{RoB67, 
Andrey_geo}. This angular distribution includes all multipoles of the 
photon field which are allowed for the given radiative transition.  In 
the particular case of the $2^+ \to 0^+$ nuclear decay, only the 
electric quadrupole transition $E2$ is allowed by the selection rules 
for which Eq.~(\ref{angular_distribution_general}) simplifies to
\begin{equation}
   \frac{d\sigma_{\rm NEEC}}{d\Omega}(\theta) = 
   \frac{\sigma_{\rm NEEC}}{4\pi} W(\theta) \,,
\end{equation}
with the angular distribution given by
\begin{equation}  \label{diff_cs}
   W(\theta) = \left(1-\frac{\sqrt{70}}{14}
   \mathcal{A}_2 P_2({\rm cos}\,\theta)-\frac{2\sqrt{14}}{7}
   \mathcal{A}_4 P_4({\rm cos}\,\theta)\right) \,. 
\end{equation}
As seen from the expression above, the angular distribution of the electric 
quadrupole transition is entirely determined by the alignment 
parameters $\mathcal{A}_2$ and $\mathcal{A}_4$ of the excited $2^+$
nuclear state and the geometrical coefficients. In the next section
we calculate these parameters and, 
hence, the angular distribution of the de--excitation photons for the 
electron capture into the $1s_{1/2}$ and $2s_{1/2}$ states of initially 
bare or He-like ${}^{174}_{70}\mathrm{Yb}$, ${}^{170}_{68}\mathrm{Er}$, 
${}^{154}_{64}\mathrm{Gd}$ and ${}^{162}_{66}\mathrm{Dy}$ ions.


\section{\label{results} Results and discussion}


In this section we present the alignment parameters and the angular 
distribution of the emitted photons that follow NEEC
into the $1s$ orbitals of bare ions and the $2s$ states of initially He-like ions.
We consider the 
even-even nuclei ${}^{174}_{70}\mathrm{Yb}$, ${}^{170}_{68}\mathrm{Er}$, 
${}^{154}_{64}\mathrm{Gd}$ and ${}^{162}_{66}\mathrm{Dy}$ for which NEEC 
total cross sections for the capture into the $K$-shell have been 
presented in Ref.~\cite{neec}. The reduced transition probability $B(E2,0\to 
2)$ for these nuclei as well as the energies of the nuclear transitions 
are taken from Ref.~\cite{Raman}. The calculation of the statistical tensors 
involves the numerical integration of $R_{L,\kappa_d,\kappa}$. For the 
continuum electron we use relativistic Coulomb-Dirac wave functions, 
applying the approximation that the nucleus is a point-like charge.
 We assume that the free electron, which is far away from the ion, is not 
sensitive to the internal structure or size of the nucleus. The radial wave functions have been calculated using the same computer routine as in Ref.~\cite{neec}, and cross-checked with the program described in Ref.~\cite{SuF05}.
The Coulomb 
phases calculated according to Eq.~(\ref{C-phase}) do not include the 
effect of the finite nuclear size, which is expected to be negligible.  For the continuum electrons recombining into the $2s$ orbital of an initially He-like ion, we use an effective nuclear charge number of $Z_{\rm eff}=Z-2$. This approximation is assumed to be sufficient for the present level of accuracy.
We consider relativistic wave functions calculated with the {\textsc 
GRASP92} package \cite{Par96} for the bound state. The finite size of 
the nucleus, i.e., its non-zero radius $R_0$, is considered in the bound wave
functions and has a sensitive effect on the inner-shell level energies of the bound electron.
The nuclear radius in calculated according to the semi-empirical formula~\cite{Soff}
\begin{equation}
R_0=(1.0793\,A^{1/3}+0.73587)\ \mathrm{fm}\ ,
\end{equation}
where $A$ is the atomic mass number.

In the case of recombination into the $2s$ orbital of an initially He-like ion,
the phase shifts of the partial continuum wave functions were calculated by 
considering the Dirac-Fock (DF) approximation of the $1s^2$ ground state 
seen by the free electron. As shown in Table~\ref{phases}, the
difference between such a calculation and 
the $Z_{\rm eff}=Z-2$ full screening approximation is very small in the 
Coulomb phase, less than 0.01~rad and has a negligible effect on the 
alignment parameters. In Table~\ref{phases} we present the Coulomb 
phases for the capture into the $1s$ orbital and $2s$ orbitals of the 
considered ions, the latter calculated using both approximations.

\begin{table*}
\caption{\label{phases} Partial wave phase shifts ($\Delta_{\kappa}$, in radian)
for capture into the $1s$ and $2s$ orbital of bare ions (Coulomb phases, third and fourth
column) and into the $1s^2 2s$ state of initially He-like ions, in different approximations. 
In the fifth column, phases for He-like ions with an effective nuclear charge
$Z_{\rm eff}=Z-2$ are given and the last column (DF) contains phases corrected for
bound electron screening in the Dirac-Fock approximation.}
\begin{center}
\begin{tabular}{l@{$\quad$}r@{$\quad$}dddd}
\hline\hline
 &   & \multicolumn{1}{c}{$1s$}& \multicolumn{3}{c}{$2s$}\\
\cline{4-6}
\multicolumn{1}{c}{${}^A_Z\mathrm{X}$} &\multicolumn{1}{l}{$\kappa$} &
\multicolumn{1}{c}{$\Delta_{\kappa}(Z_{\rm eff}=Z)$} &
\multicolumn{1}{c}{$\Delta_{\kappa}(Z_{\rm eff}=Z)$} &
\multicolumn{1}{c}{$\Delta_{\kappa}(Z_{\rm eff}=Z-2)$} &
\multicolumn{1}{c}{$\Delta_{\kappa}({\rm DF})$} \\ 
\hline
${}^{154}_{64}\mathrm{Gd}$ &  2 &  2.313 &  2.507 &  2.524 &  2.529 \\
                           & -3 & -0.913 & -0.731 & -0.709 & -0.706 \\
${}^{162}_{66}\mathrm{Dy}$ &  2 &  0.747 &  2.246 &  2.270 &  2.273 \\
                           & -3 & -2.461 & -0.983 & -0.954 & -0.952 \\
${}^{170}_{68}\mathrm{Er}$ &  2 &  0.805 &  2.254 &  2.277 &  2.280 \\
                           & -3 & -2.408 & -0.980 & -0.952 & -0.951 \\
${}^{174}_{70}\mathrm{Yb}$ &  2 &  5.293 &  2.205 &  2.229 &  2.231 \\
                           & -3 &  2.079 & -1.033 & -1.004 & -1.003 \\
\hline
\hline
\end{tabular}
\end{center}
\end{table*}

\begin{figure}
\begin{center}
\includegraphics[width=0.45\textwidth]{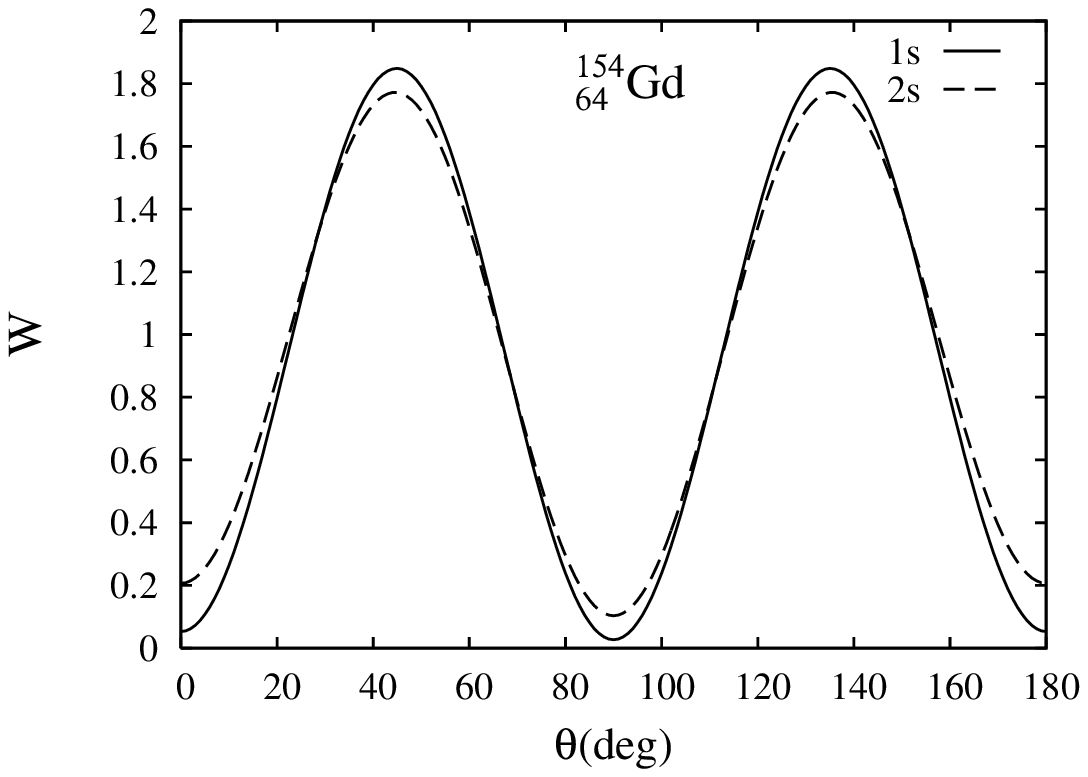}\includegraphics[width=0.45\textwidth]{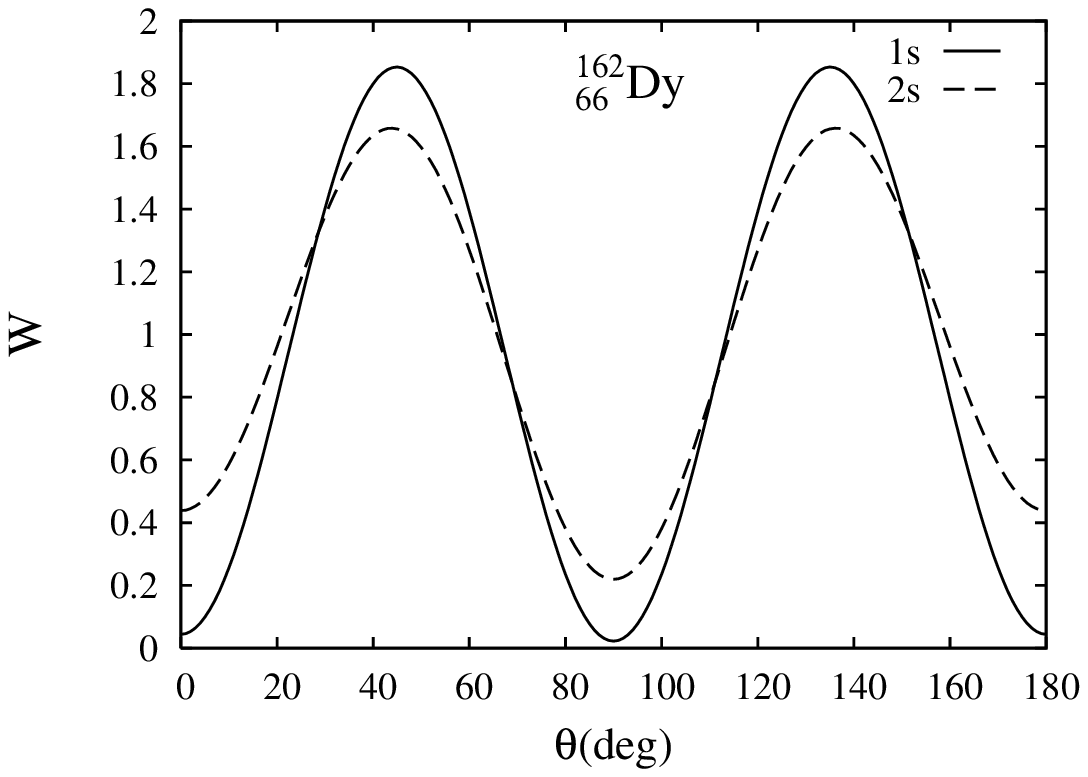}
\includegraphics[width=0.45\textwidth]{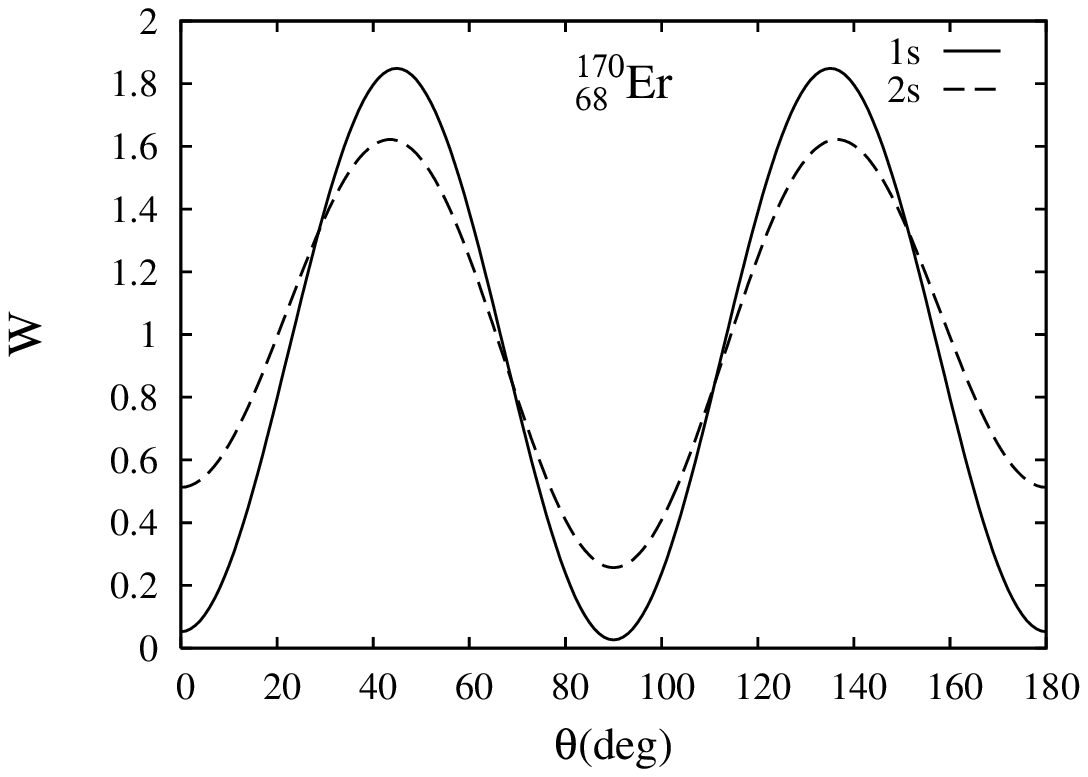}\includegraphics[width=0.45\textwidth]{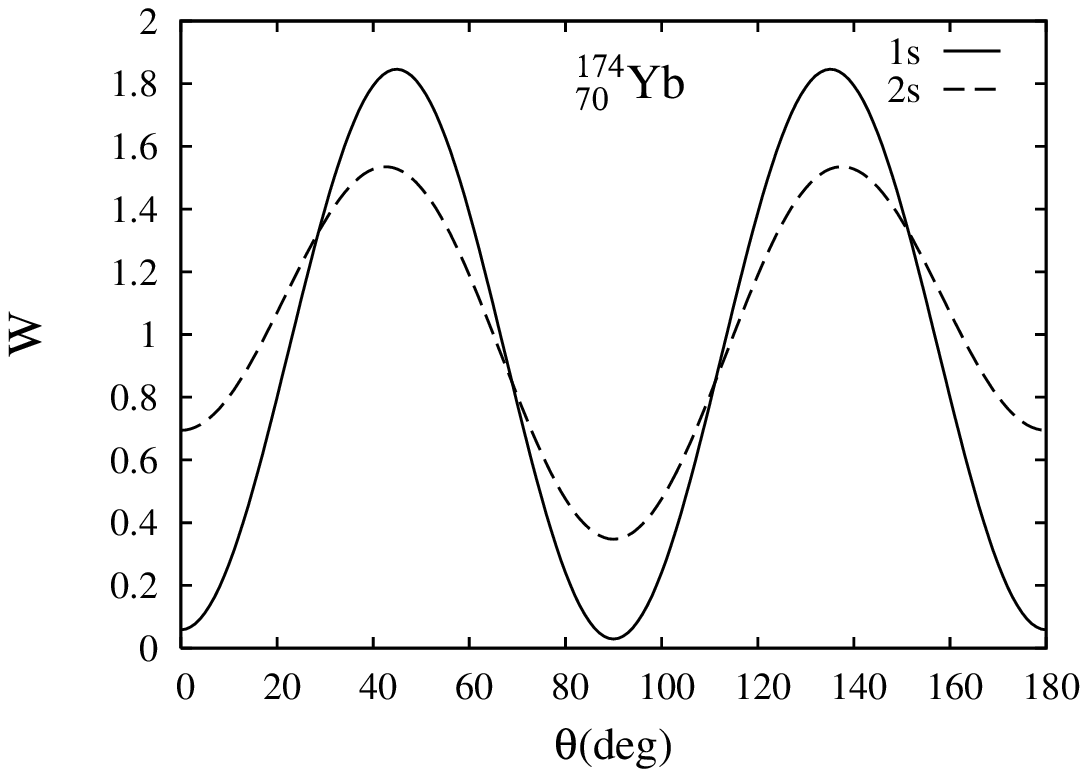}
\caption{\label{wees} Angular distribution $W(\theta)$ of photons emitted 
in the radiative nuclear decay of the $2^+$ excited state following NEEC 
for the elements ${}^{154}_{64}\mathrm{Gd}$, ${}^{164}_{66}\mathrm{Dy}$, 
${}^{170}_{68}\mathrm{Er}$, and ${}^{174}_{70}\mathrm{Yb}$, respectively. 
The cases of recombination into the $1s$ state of initially bare ions 
(solid lines) and into the $2s$ orbital of initially He-like ions (dashed lines) are 
presented.}
\end{center}
\end{figure}

The values of the alignment parameters $\mathcal{A}_2$ and 
$\mathcal{A}_4$ of the $2^+$ excited nuclear states are presented in 
Table \ref{align}. Capture into the $1s$ as well as the $2s$ orbitals are 
considered. The alignment of the excited nuclear state characterized by 
these parameters gives in the second step of NEEC the angular distribution
of the emitted radiation. In Figure~\ref{wees} 
we present the angular distribution $W(\theta)$ given in 
Eq.~(\ref{diff_cs}) for the capture into the $1s$ and $2s$ shells of 
${}^{154}_{64}\mathrm{Gd}$, ${}^{162}_{66}\mathrm{Dy}$, 
${}^{170}_{68}\mathrm{Er}$ and ${}^{174}_{70}\mathrm{Yb}$, respectively. The 
angular patterns are similar for all the four ions, as they all involve 
$E2$ transitions of nuclei with near atomic and mass numbers. Both 
radiations following the capture into the $1s$ and $2s$ orbitals of the 
ion present maxima at $\theta=45^{\circ}$ and $\theta=135^{\circ}$, 
a pattern which significantly differs from that of RR~\cite{Andrey}.
While for the capture into the $1s$ orbitals the radiation intensity
emitted at $\theta=0^{\circ},90^{\circ},180^{\circ}$ appears to be
negligible, the pattern for the capture into $2s$ displays larger
minimum values at these angles. In contrast to NEEC, 
the radiative recombination of the free electron is dominated by the $E1$ 
transition and has an angular distribution (roughly given by a
${\rm sin}^2\theta$ function) with a maximum at $\theta=78^{\circ}$ in the center-of-mass frame.

\begin{table*}
\caption{\label{align} Alignment parameters for the $2^+$ excited 
nuclear state formed by NEEC. We denote the capture orbital by $nl_j$ and 
$E_{c}$ is the energy of the continuum electron at the resonance.}
\begin{center}
\begin{tabular}{lr@{$\quad$}c@{$\quad$}c @{$\quad$}c}
\hline\hline
${}^A_Z\mathrm{X}$ & $E_{c}$(keV) & $nl_j$ &$\mathcal{A}_2$&$\mathcal{A}_4$ \\ 
\hline
${}^{154}_{64}\mathrm{Gd}$ & 64.005  & $1s_{1/2}$ & -1.183 & 1.547\\
${}^{164}_{66}\mathrm{Dy}$ & 10.318  & $1s_{1/2}$ & -1.185 & 1.557\\
${}^{170}_{68}\mathrm{Er}$ & 11.350  & $1s_{1/2}$ & -1.183 & 1.548\\
${}^{174}_{70}\mathrm{Yb}$ &  4.897  & $1s_{1/2}$ & -1.181 & 1.542\\
${}^{154}_{64}\mathrm{Gd}$ & 108.847 & $2s_{1/2}$ & -1.146 & 1.383\\
${}^{164}_{66}\mathrm{Dy}$ &  58.164 & $2s_{1/2}$ & -1.090 & 1.135\\
${}^{170}_{68}\mathrm{Er}$ &  62.317 & $2s_{1/2}$ & -1.073 & 1.056\\
${}^{174}_{70}\mathrm{Yb}$ &  59.106 & $2s_{1/2}$ & -1.029 & 0.861\\
\hline\hline
\end{tabular}
\end{center}
\end{table*}

\begin{figure}
\begin{center}
\includegraphics[width=0.6\textwidth]{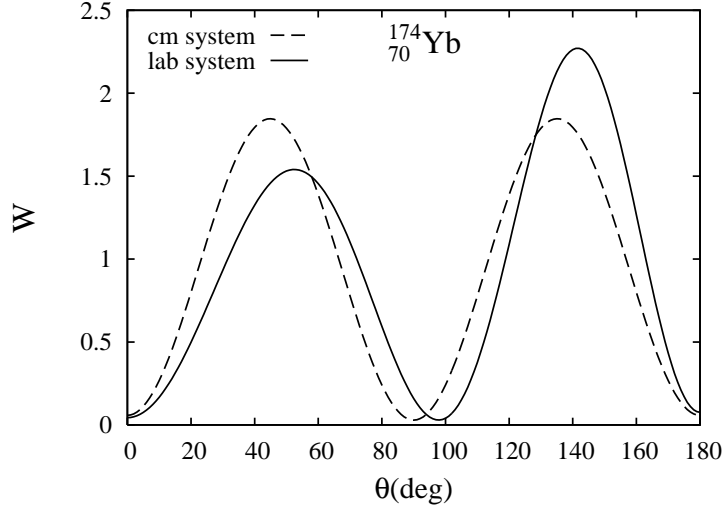}
\caption{\label{yb_lo} Angular distribution of the photons with respect 
to the laboratory (lab) and center-of-mass (cm) systems for the case of NEEC into 
the $1s$ orbital of ${\rm Yb}^{70+}$. }
\end{center}
\end{figure}

As the RR and NEEC angular distributions of the emitted photons have 
maxima at different values of $\theta$, we calculate the ratio between 
the two angular differential cross sections at different emission 
angles,
\begin{equation}
R(E)=\left.\left(\frac{d\sigma_{\rm NEEC}}{d\Omega}(E,\theta) 
\Big/\frac{d\sigma_{\rm RR}}{d\Omega}(E,\theta)\right)\right|_{\theta=\theta_{max}}\ 
\label{r_ang}
\end{equation}
for the case of electron capture into bare ytterbium. We consider here 
the angles $\theta_{max}=45^{\circ},78^{\circ},135^{\circ}$ that 
correspond to the maxima of NEEC and RR radiation angular distributions. The NEEC 
total cross section is convoluted with the energy distribution of the 
continuum electrons assuming a Gaussian width parameter of 0.1~eV. 
The RR angular differential cross section is calculated within the 
framework of Dirac's relativistic equation and by taking into account 
the higher (non--dipole) terms in the expansion of the electron--photon 
interaction \cite{Andrey,Eichler,FrI05}. We assume that the RR and NEEC 
alignment parameters are constant on the studied energy interval of 
approximately 1~eV.

We envisage the scenario of a possible NEEC experiment in a storage 
ring, in which the radiation is emitted by the nucleus of the 
$\mathrm{Yb}^{69+}$ ion moving relativistically with respect to the 
laboratory frame. A Lorentz transformation of the NEEC and RR angular 
differential cross sections in the center-of-mass frame is therefore 
required in order to obtain the quantities in the laboratory system. The 
angular differential cross section in the laboratory system can be 
written as \cite{Eichler}
\begin{equation}
\frac{d\sigma(\theta)}{d\Omega} = 
\frac{1}{\gamma^2(1-\beta\cos\theta)^2} \frac{d\sigma'(\theta')}{d\Omega'}\,,
\label{lo_transf}
\end{equation}
where $d\sigma'(\theta')/d\Omega'$ is the differential cross section in 
the center-of-mass system, denoted until now by the unprimed symbols. In our case the reduced velocity $\beta$ is 
0.138 and the Lorentz factor is $\gamma=1.009$. The angle of the photons 
in the laboratory frame $\theta$ is related to the one in the ion-fixed 
frame $\theta'$ by
\begin{equation}
\cos\theta'=\frac{\cos\theta-\beta}{1-\beta\cos\theta}\ .
\end{equation} 
As the system possesses azimuthal symmetry, $\varphi'=\varphi$. The 
angular distribution of the photons with respect to the laboratory 
system for the case of ytterbium is presented in Fig.~\ref{yb_lo}.

\begin{figure}
\begin{center}
\includegraphics[width=0.6\textwidth]{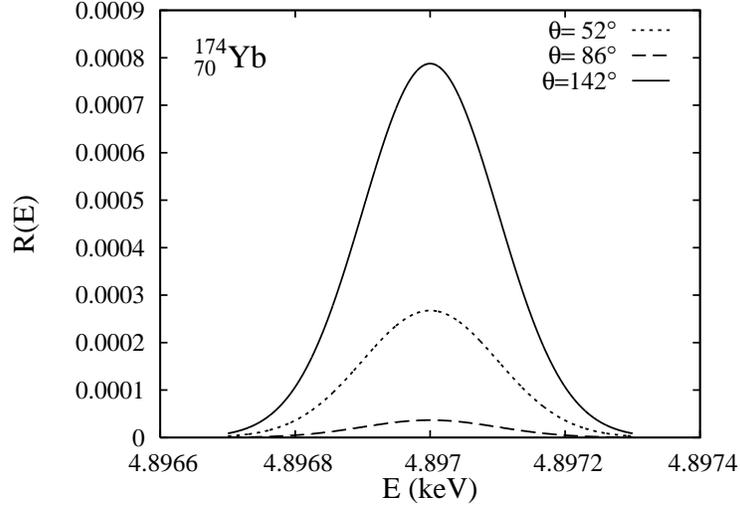}
\caption{\label{ang_ratio} The NEEC and RR angular differential cross 
sections ratio for the case of ${}^{174}_{70}\mathrm{Yb}$ as a function of 
the continuum electron energy for three different photon emission 
angles. The NEEC total cross section was convoluted with a Gaussian 
electron energy distribution with the width parameter of 0.1~eV for a better visual representation of the data.}
\end{center}
\end{figure}

In Fig.~\ref{ang_ratio} we present the ratio in Eq.~(\ref{r_ang}) as a 
function of the continuum electron energy for the three values of the 
photon emission angle $\theta$ in the laboratory frame, for which $d\sigma_{\scriptscriptstyle \rm NEEC}
(\theta)/d\Omega$ or $d\sigma_{\scriptscriptstyle \rm RR}(\theta)/d\Omega$
have a maximum. In the laboratory frame, the NEEC angular distribution has
maxima at $\theta=52^{\circ}$ and $142^{\circ}$, while in the case of RR
the peak is at approximately $\theta=86^{\circ}$. The ratio of the NEEC and
RR angular differential cross sections is more than one order of magnitude 
larger for $\theta=52^{\circ}$ and $142^{\circ}$ than in the case of 
$\theta=86^{\circ}$. If the photons emitted perpendicular to the 
direction of the incoming electron are measured in an experiment, it is 
most likely that only the RR background will be detected, as the NEEC 
differential cross section presents a minimum at $\theta=98^{\circ}$.
For the emission angles $\theta=52^{\circ}$ and $\theta=142^{\circ}$
the RR contribution at the resonance electron energy $E_c$
is considerably lower in comparison to the total
cross sections of the recombination process,
\begin{eqnarray}
\left.\left(\frac{d\sigma_{\rm NEEC}}{d\Omega}(E_c,\theta) 
\Big/\frac{d\sigma_{\rm RR}}{d\Omega}(E_c,\theta)\right)\right|_{\theta=52^{\circ}} & \simeq &
1.5\ \frac{\sigma_{\rm NEEC}(E_c)}{\sigma_{\rm RR}(E_c)}\,, \\
\left.\left(\frac{d\sigma_{\rm NEEC}}{d\Omega}(E_c,\theta)
\Big/\frac{d\sigma_{\rm RR}}{d\Omega}(E_c,\theta)\right)\right|_{\theta=142^{\circ}} & \simeq &
4.4\ \frac{\sigma_{\rm NEEC}(E_c)}{\sigma_{\rm RR}(E_c)}\,.
\end{eqnarray}

As the ratio of the NEEC and RR angular differential cross sections is 
rather small, the experimental observation of the NEEC signature is challenging.
Nevertheless, knowing the angular 
pattern of NEEC is important as it provides a means of suppressing the RR 
background. Storage ring experiments focused on detecting the photons 
emitted in photo-recombination have the best chances to observe the NEEC resonance
at an angle of $\theta=142^{\circ}$ for the case of $E2$ 
transitions of the ${}^{174}_{70}\mathrm{Yb}$ nucleus.

\section{\label{summary} Summary}


In this work we consider the alignment of nuclear states formed in the process of
nuclear excitation by electron capture. We calculate alignment parameters
and geometric factors which determine the angular distribution of the 
subsequently emitted gamma photons. Such distribution functions are presented
for a range of heavy elements with $E2$ nuclear de-excitation transitions. As the
emission pattern of nuclear gamma photons is found to be substantially different from 
the emission characteristics of the background process of radiative recombination,
our findings may help experimental attempts to discern NEEC from the competing process.
E.g., in the case of the ${}^{174}_{70}\mathrm{Yb}$ nucleus with an $E2$ transition,
the best chance to observe the NEEC resonance is at an angle of $\theta=142^{\circ}$.
This emphasizes the importance of measuring photon angle-resolved cross sections.
Furthermore, the knowledge of the maxima and minima of the angular distribution of gamma
radiation following NEEC is of general interest for any measurement aiming to 
observe NEEC.


\begin{acknowledgments}

AP and ZH appreciate valuable discussions with Prof. Werner Scheid. 
AP and UDJ acknowledge support from the Deutsche Forschungsgemeinschaft (DFG).

\end{acknowledgments}

\bibliography{ang}

\end{document}